\documentclass[aps,prl,twocolumn]{revtex4}
\usepackage{amssymb}
\usepackage{graphicx}
\usepackage{hyperref}

\begin{document}

\title{Stretched Exponential Relaxation Arising from a Continuous Sum of
Exponential Decays}
\author{D. C. Johnston}
\affiliation{Ames Laboratory and Department of Physics and Astronomy, Iowa State
University, Ames, Iowa 50011}

\date{\today}

\begin{abstract}

Stretched exponential relaxation of a quantity $n$ versus time $t$ according to
$n = n_0 \exp[-(\lambda^* t)^\beta]$ is ubiquitous in many
research fields, where 
$\lambda^*$ is a characteristic relaxation rate and the stretching exponent
$\beta$ is in the range $0 < \beta < 1$.   Here we consider systems in which
the stretched exponential relaxation arises from the global relaxation of
a system containing independently exponentially relaxing species with a
probability distribution $P(\lambda/\lambda^*,\beta)$ of relaxation rates
$\lambda$.  We study the properties of $P(\lambda/\lambda^*,\beta)$ and their
dependence on $\beta$.  Physical interpretations of $\lambda^*$ and $\beta$,
derived from consideration of $P(\lambda/\lambda^*,\beta)$ and its moments, are
discussed.

\end{abstract}

\pacs{75.40.Gb, 82.56.Na, 75.10.Nr, 71.55.Jv}

\maketitle

\section{\label{intro}Introduction}

\begin{figure}[t]
\includegraphics[width=3in]{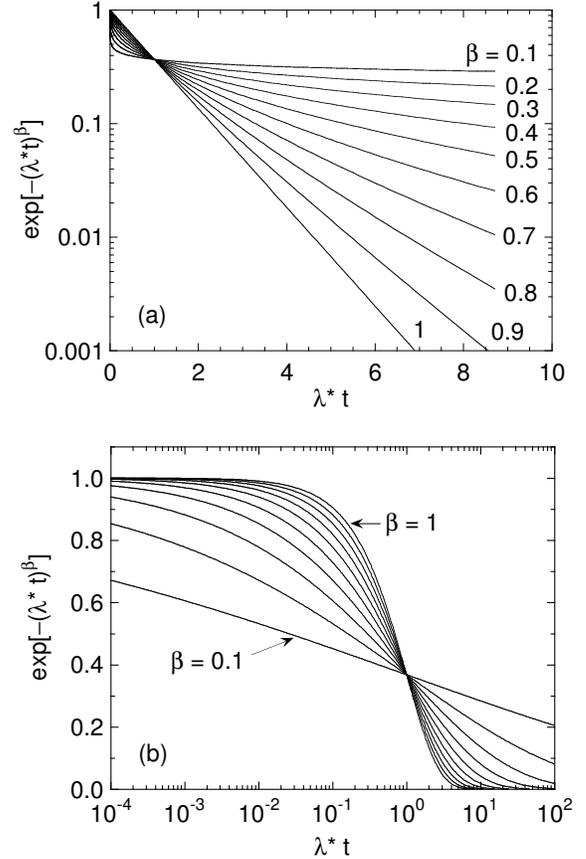}
\caption{Log-linear (a) and linear-log (b) plots of the stretched exponential
function in Eq.\ (\protect\ref{EqStretch}) versus $\lambda^* t$ for $\beta$ values
from 0.1 to 1 in 0.1 increments.}
\label{FigStretchedExponentials}
\end{figure}

The stretched exponential relaxation function describes the time $t$ dependence of
a relaxing quantity $n$, according to
\begin{equation} 
n = n_0 \exp[-(\lambda^* t)^\beta]\ ,
\label{EqStretch}
\end{equation}
where $n_0 \equiv n(t = 0)$, $\lambda^*$ is
a characteristic relaxation rate, and the stretching exponent $\beta$ is in the
range $0 < \beta < 1$.  This behavior has been observed for a wide variety of
physical quantities in many different systems and research
areas.\cite{Phillips1996}  Log-linear and linear-log plots of the stretched
exponential function versus $\lambda^* t$, which are the two most common ways of
plotting the stretched exponential function, are shown in Figs.\
\ref{FigStretchedExponentials}(a) and \ref{FigStretchedExponentials}(b),
respectively, for $\beta$ values from 0.1 to 1 in 0.1 increments.  All of the 
plots cross at a time $t_{1/e} =
1/\lambda^*$ at which the stretched exponential function has the value $e^{-1}$
for all $\beta$.  A pure
exponential decay, corresponding to $\beta = 1$, plots as a straight line in
Fig.\ \ref{FigStretchedExponentials}(a).  As
$\beta$ decreases from unity, more and more pronounced positive curvature is
evident at small $t$.  At small times, a Taylor series expansion of Eq.\
(\ref{EqStretch}) for $\lambda^* t \ll 1$ gives 
\begin{equation}
{n\over n_0}(t \to 0) \approx 1- (\lambda^* t)^\beta\ .  
\label{Eqtto0}
\end{equation}
Thus the
stretched exponential function with $0 < \beta < 1$ is singular at $t = 0$, with
an infinitely negative slope there.  

A natural and commonly used interpretation of an observed stretched exponential
relaxation is in terms of the global relaxation of a system containing many
independently relaxing species, each of which decays exponentially in time with a
specific fixed relaxation rate $\lambda$.  Then one can write the stretched
exponential function as a continuous sum of pure exponential decays, with a
particular probability distribution $P$ of $\lambda$ values for a given value of
$\beta$.  In such a probability distribution, one must normalize $\lambda$ to the
characteristic relaxation rate $\lambda^*$ appearing in the stretched
exponential function (\ref{EqStretch}).  Hence for such systems one can write the
stretched exponential function in Eq.\ (\ref{EqStretch}) as
\begin{equation}  
e^{-(\lambda^* t)^\beta} = \int_0^\infty P(s,\beta)e^{-s\lambda^* t}ds\ ,
\label{EqProb}
\end{equation}
where
\[
s\equiv{\lambda\over\lambda^*}\ .
\]
Since $P(s,\beta)$ is a probability density, one has $\int_0^\infty P(s,\beta) ds
= 1$.

A stretched exponential form of the $^7$Li nuclear spin-lattice relaxation
following saturation was recently observed\cite{Trinkl2000,Kaps2001,Johnston2005}
in $^7$Li NMR experiments on the heavy fermion\cite{Kondo1997} compound ${\rm
LiV_2O_4}$ containing magnetic defects.\cite{Johnston2005}  In our attempts to
understand the origin of this nonexponential relaxation, we considered the above
model of independently relaxing nuclear spins and found the published
information on $P(s,\beta)$
(e.g., Refs.\ \onlinecite{Pollard1946, Williams1970, Lindsey1980,Helfand1983,%
Montroll1984, Weron1986, Svare2000, Hetman2003}) 
to be insufficient for our purposes, particularly with regard to the 
systematic behaviors of $P(s)$ versus $\beta$ and to the physical significances of the parameters $\lambda^*$ and $\beta$ in the
stretched exponential function (\ref{EqStretch}).  Since in the case of
independently relaxing species the essential physics of the system resides
in $P(s,\beta)$ rather than in the relaxation function (\ref{EqStretch}) itself,
we considered it important to further study $P(s,\beta)$.  Here we report the
results,  some of which were briefly mentioned in Ref.\
\onlinecite{Johnston2005}.

\section{Results}

\subsection{Probability density}

For $\beta = 1$, the stretched exponential function (\ref{EqStretch}) is a pure
exponential with relaxation rate $\lambda = \lambda^*$ and hence the probability
density $P(s,1)$ in Eq.\ (\ref{EqProb}) is a Dirac $\delta$ function at $s = 1$. 
For general $\beta$, from Eq.\ (\ref{EqProb}) one sees that $P(s,\beta)$ is the
inverse Laplace transform of the stretched exponential, given by 
\begin{equation} 
P(s,\beta) = {1\over{2\pi}i}\int_{-i\infty}^{i\infty}e^{-x^\beta}e^{sx}dx\ .
\label{EqPLaplace0}
\end{equation} 
The change of variables $u = -ix$ allows one to write $P(s,\beta)$ as the Fourier
transform
\begin{equation} 
P(s,\beta) = {1\over{2\pi}}\int_{-\infty}^{\infty}e^{-(iu)^\beta}e^{isu}du\ .
\label{EqPLaplace1}
\end{equation} 
One can also express $P(s,\beta)$ as\cite{Pollard1946,Lindsey1980,Weron1986}
\begin{equation}
P(s,\beta) = {1\over \pi}\sum_{n=1}^\infty {(-1)^{n+1} \Gamma(n\beta + 1)\over n!
s^{n\beta+1}} \sin(n\pi\beta)\ ,
\label{EqPLaplace3}
\end{equation}
where $\Gamma(z)$ is the Gamma (factorial) function.  Using Mathematica 4.0,
we have obtained from Eqs.~(\ref{EqPLaplace1}) and (\ref{EqPLaplace3})
closed analytic solutions for $P(s,\beta)$ for many rational values of $\beta$. 
The simplest expressions are obtained for $\beta = 1/3$, 1/2, and 2/3, for which
solutions have also been given in Refs.\ \onlinecite{Helfand1983} and
\onlinecite{Montroll1984}.  For
$\beta = 1/3$, we find the two alternative expressions
\begin{eqnarray}
P\big(s,{1\over 3}\big) &=& 3z^4 {\rm Ai}(z)\nonumber\\
&=& {1\over 3\pi s^{3/2}} K_{1/3}\Big({2\over \sqrt{27s}}\Big)\ ,
\end{eqnarray}
where ${\rm Ai}(z)$ is the Airy function with $z=(3s)^{-1/3}$ and $K_n(x)$ is the
modified Bessel function of the second kind.  For $\beta = 1/2$, we get
\begin{equation}
P\big(s,{1\over 2}\big) = {1\over \sqrt{4\pi s^3}} \exp\big(-{1\over 4s}\big)\ ,
\label{EqP12}
\end{equation}
which is also listed in tables of Laplace transforms.\cite{Abramowitz1972} For
$\beta = 2/3$, we obtain
\begin{equation}
P\big(s,{2\over 3}\big) = 6z^{7/4}\exp\big(-{2\over 27s^2}\big)[{\rm Ai}(z) -
{1\over\sqrt{z}} {\rm Ai}^\prime(z)]\ , 
\end{equation}
where $z = (3s)^{-4/3}$.

In general, the analytic results for rational $\beta = n/m$
values are expressed by Mathematica in terms of generalized hypergeometric
functions $_pF_q(\mathbf{a};\mathbf{b};z)$.  For example, for $\beta =
2/3$, $P(s,\beta)$ can also be expressed as
\begin{eqnarray}
P\Big(s,{2\over 3}\Big) &=& {1\over \sqrt{3} \pi
s^{7/3}}\Big[s^{2/3}\Gamma\Big({2\over 3}\Big)\,_1F_1\Big({5\over 6},{2\over
3},z\Big)\nonumber\\ &+& \Gamma\Big({4\over
3}\Big)\,_1F_1\Big({7\over 6},{4\over 3},z\Big)\Big]\ ,
\end{eqnarray}
with $z = -4/(27 s^2)$.  The expressions become progressively
longer as the integer denominator $m$ increases.  For example,  
the expression for $P(s,\beta = 4/5)$ contains four additive
$_pF_q(\mathbf{a}; \mathbf{b}; z)$ terms with $p = q = 3$ and $z = -256/(3125
s^4)$, with $1/s^{1+n\beta}$ multiplicative prefactors where $n = 1$--4,
respectively.

The probability density $P(s,\beta)$ in Eq.\ (\ref{EqProb}) is real by
definition.  By finding the real and imaginary parts of the integral on the right
side of Eq.\ (\ref{EqPLaplace1}), one can explicitly show that the imaginary part
is identically zero.  The real part, which is
$P(s,\beta)$, is found to be
\begin{equation}
P(s,\beta)={1\over \pi}\int_0^\infty e^{-u^\beta {\rm cos}(\pi\beta/2)}{\rm
cos}[s u - u^\beta{\rm sin}(\pi\beta/2)]du\ .
\label{EqPLaplace2}
\end{equation}
This integral can be evaluated numerically for arbitrary $\beta$ using
Mathematica.  Similar expressions were obtained in Refs.\
\onlinecite{Lindsey1980} and \onlinecite{Weron1986}.

\begin{figure}[t]
\includegraphics[width=3in]{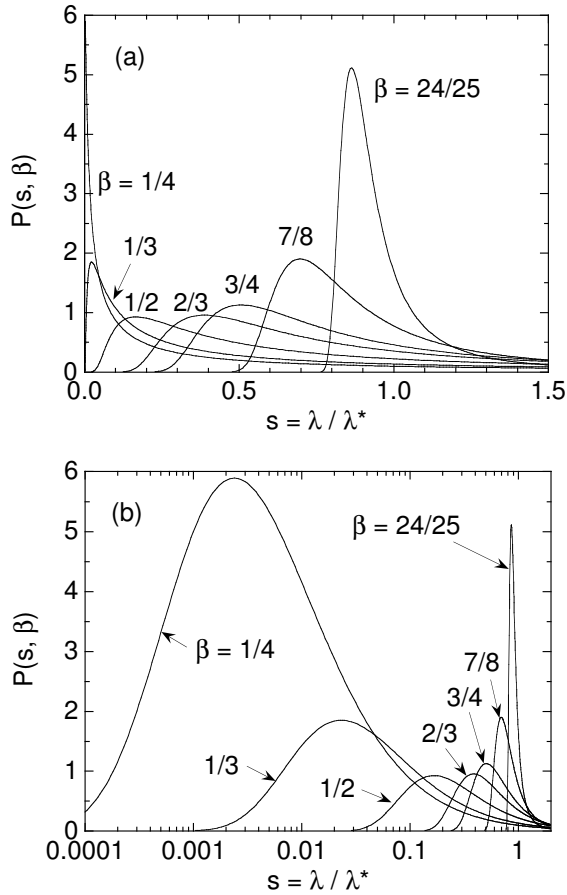}
\caption{Linear (a) and semilog (b) plots of the probability density
$P(s,\beta)$ for the relaxation rate in Eq.\ (\protect\ref{EqProb}) versus
normalized relaxation rate $s = \lambda/\lambda^*$ for several rational values of
$\beta$.}
\label{FigP(s,beta)}
\end{figure}

The evolution of $P(s,\beta)$ versus $s$ at fixed $\beta$ for several
rational values of $\beta$, from our analytic results, is plotted in Fig.\
\ref{FigP(s,beta)}(a) on linear scales and in Fig.\
\ref{FigP(s,beta)}(b) on semilog scales.  The same type of plots as in Fig.\
\ref{FigP(s,beta)}(a) were previously given in Fig.\ 1 of Ref.\
\onlinecite{Hetman2003}, and related plots were given in Refs.\
\onlinecite{Lindsey1980} and \onlinecite{Svare2000}.  From Fig.\
\ref{FigP(s,beta)}(a), a
$\beta < 1$ causes the infinitely high and narrow Dirac $\delta$ function
probability distribution for $\beta = 1$ at $s = 1$ to broaden.  $P(s)$ becomes
highly asymmetric, and the peak in $P(s)$ becomes finite and moves towards slower
rates which is compensated by a long tail to faster rates.  

\begin{figure}[t]
\includegraphics[width=3in]{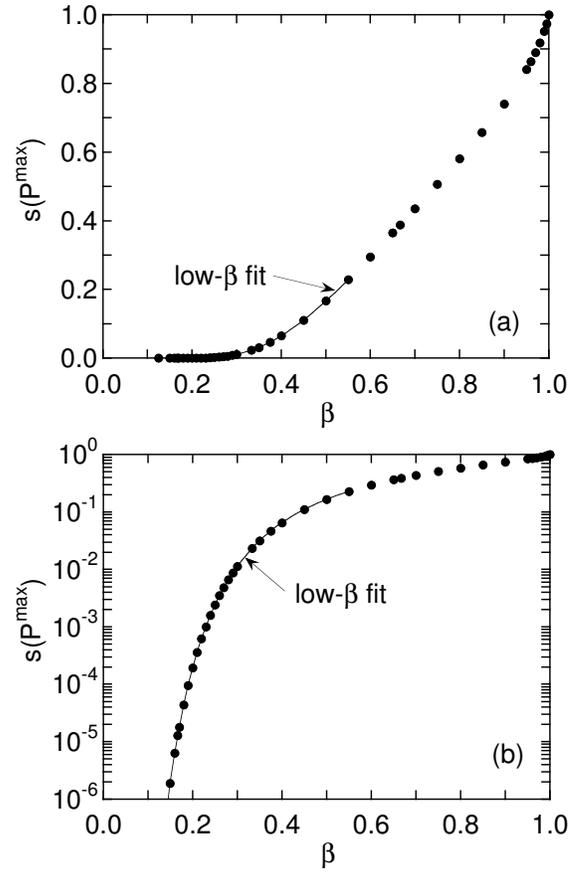}
\caption{Linear (a) and semilog (b) plots of the value $s(P^{\rm max})$ of $s$
at which the probability density $P(s)$ in Eq.\ (\protect\ref{EqProb}) (see
Fig.\ \protect\ref{FigP(s,beta)}) is maximum, plotted versus the stretching
exponent $\beta$.  The solid curve in (a) and (b) is an exponential fit to the
data for $\beta \leq 0.5$ as given in Eq.\ (\protect\ref{EqPfit}).  The position
of the peak can be taken to be the intrinsic small-$s$ cutoff to $P(s)$ 
(see text).}
\label{Figs^max(beta)}
\end{figure}

The value of $s$ at which $P(s,\beta)$ is maximum for fixed $\beta$, $s(P^{\rm
max})$, is plotted versus $\beta$ on linear and semilog scales in Figs.\
\ref{Figs^max(beta)}(a) and (b), respectively.  We find that $s(P^{\rm max})$
decreases with decreasing $\beta$ and approaches zero exponentially for
$\beta\lesssim 0.5$.  A fit to the $s(P^{\rm max})$ versus $\beta$ data in the
range $1/8 \leq \beta \leq 1/2$ yielded
\begin{equation}
s(P^{\rm max}) \approx {2.56\over \beta^{0.6}}
\exp\big(-{1.27\over\beta^{1.31}}\big) \ .
\label{EqPfit}
\end{equation}
Plots of this expression are shown as the solid curves in Figs.\
\ref{Figs^max(beta)}(a) and \ref{Figs^max(beta)}(b). 

For small $s$, there is an exponential decrease in $P(s,\beta)$ with decreasing
$s$ that is given in the limit $s\to 0$ (i.e., below
the peak) by\cite{Helfand1983,Weron1986}
\begin{equation}
P(s\to 0,\beta)={a\over s^{1+\beta/[
2(1-\beta)]}}\exp\big(-{b\over s^{\beta/(1-\beta)}}\big) \ ,
\label{EqPsto0}
\end{equation}
where
\[
a = {\beta^{1 + \beta/[2(1-\beta)]}\over \sqrt{2\pi\beta(1-\beta)}}
\]
and
\[
b = (1-\beta)\beta^{\beta/(1-\beta)}\ .
\]
From Eq.\ (\ref{EqP12}) for $\beta = 1/2$ one gets directly that
$a = 1/\sqrt{4\pi}$ and $b = 1/4$, in agreement with these results.  Thus the
probability distribution $P(s,\beta)$ contains an intrinsic low-$s$ cutoff that
decreases with decreasing $\beta$.  Note that the peak position of $P(s,\beta)$
in Eq.\ (\ref{EqPfit}) has the same form as the $s\to 0$ behavior in Eq.\
(\ref{EqPsto0}).  We take the value of the low-$s$ cutoff to be the position
of the peak in $P(s,\beta)$ that is plotted in Fig.\ \ref{Figs^max(beta)} and
described by Eq.\ (\ref{EqPfit}) for $\beta \lesssim 1/2$.

In the opposite limit of large $s \gg 1$, in Eq.\ (\ref{EqPLaplace3}) one retains
only the first term in the power series, giving
\begin{equation}
P(s\rightarrow \infty,\beta)= {c\over s^{1+\beta}}\ , 
\label{EqPtoInfty}
\end{equation}
where
\begin{equation}
c = {\Gamma(\beta+1)\over \pi} \sin(\pi\beta)\ .
\label{EqCoeff}
\end{equation}
For $0 < \beta < 1$, one has that $\Gamma(\beta + 1)\sim 1$.

\begin{figure}[t]
\includegraphics[width=3in]{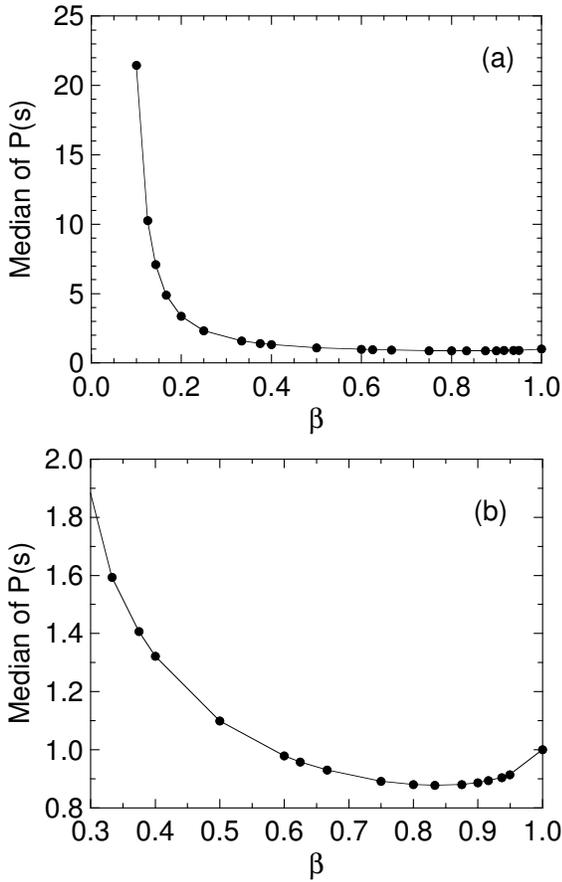}
\caption{(a) Median of the probability distribution $P(s)$
versus $\beta$.  (b) Expanded plot of the data in (a) for $0.3\leq \beta \leq 1$.
 The lines are guides to the eye.}
\label{FigP(s)Median}
\end{figure}

\begin{figure}[t]
\includegraphics[width=3in]{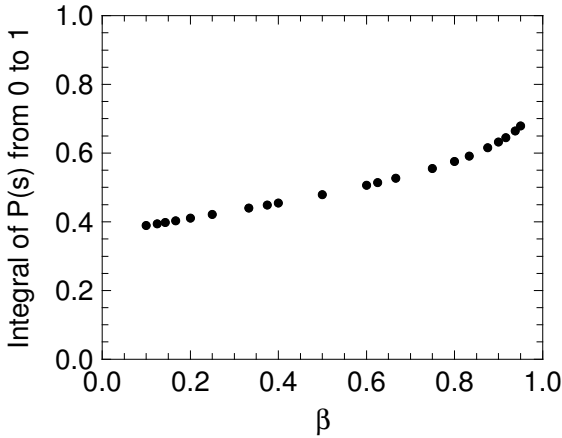}
\caption{The integral $\int_0^1P(s,\beta)ds$ versus $\beta$.}
\label{FigPIntegral(0to1)}
\end{figure}

\begin{figure}[t]
\includegraphics[width=3in]{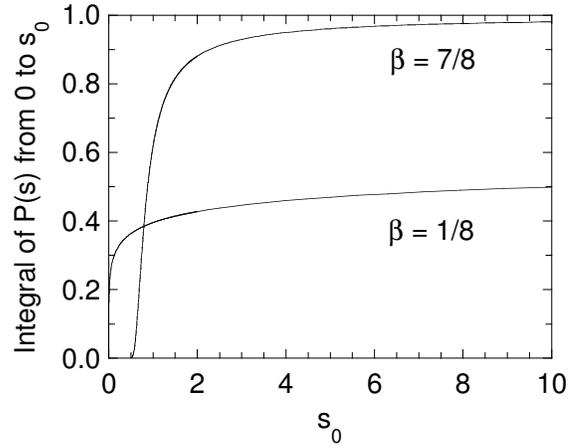}
\caption{Probability that $s$ is less than the value $s_0$, given by the integral
$\int_0^{s_0} P(s,\beta)ds$, for $\beta = 1/8$ and 7/8.}
\label{FigP_Integral}
\end{figure}

The median $s_{\rm median}$ of the probability distribution $P(s)$ has not been
discussed before in the literature to our knowledge.  It is defined to be the
value of $s$ for which it is equally likely for $s$ to be less than
$s_{\rm median}$ as it is to be greater.  It is calculated by solving the
expression $\int_0^{s_{\rm median}}P(s,\beta)ds = 1/2$ for a fixed value of
$\beta$.  The $s_{\rm median}$ is plotted versus $\beta$ in Fig.\
\ref{FigP(s)Median}(a), and an expanded plot of the data for $0.3 \leq \beta \leq 1$ is shown in Fig.\
\ref{FigP(s)Median}(b).  One sees from Fig.\  \ref{FigP(s)Median} that with
decreasing $\beta$, $s_{\rm median}$ remains nearly equal to unity from $\beta
= 1$ down to about $\beta = 0.5$, below which $s_{\rm median}$ begins to increase
dramatically.  A related quantity is the integral
$\int_0^1P(s,\beta)ds$, which measures the probability that $s$ is less than or
equal to unity.  If the median is at $s = 1$, then the integral should equal
1/2.  The integral is plotted versus $\beta$ in Fig.\
\ref{FigPIntegral(0to1)}.  In the range $0.1 \leq \beta \leq 0.9$, the integral
is equal to 1/2 to within $\pm 0.1$.  The reason for the difference between
the $\beta$ dependences of the two quantities is apparent from plots for 
$\beta = 1/8$ and 7/8 in Fig.\ \ref{FigP_Integral} of the
probability that $s$ is less than a value $s_0$, given by the integral
$\int_0^{s_0} P(s,\beta)ds$.  One sees from the figure how it happens that for
$\beta = 1/8$, $s_{\rm median} > 2\int_0^1P(s,\beta)ds$, whereas for $\beta =
7/8$, $s_{\rm median} \lesssim 2\int_0^1P(s,\beta)ds$.

\subsection{Moments of $P(s,\beta)$}

The moments of the probability distribution are defined by $(s^n)_{\rm ave} =
\int_0^\infty s^nP(s,\beta)ds$.  From Eq.\ (\ref{EqPtoInfty}), the $n = 1$
moment, which is the average $s_{\rm ave} =(\lambda/\lambda^*)_{\rm ave} =
\lambda_{\rm ave}/\lambda^*$, is infinite for all $\beta$ with $0 < \beta < 1$.  
Hence the average relaxation rate $\lambda_{\rm ave}$ is infinite.  Indeed all
positive moments such as also
$(s^2)_{\rm ave}$ are infinite.  Very fast (infinite) relaxation rate components
are required to produce the infinitely negative slope for any $\beta < 1$ in the
stretched exponential function (\ref{EqStretch}) as $t
\rightarrow 0$ as described in Eq.\ (\ref{Eqtto0}) and seen in Fig.\
\ref{FigStretchedExponentials}(a).  A cutoff to
$P(s,\beta)$ at large $s$ is required to obtain a finite initial slope of $n(t)$
in Eq.\ (\ref{EqStretch}) and finite averages
$(s^n)_{\rm ave}$.  Every real system must have a high-$s$ cutoff to $P(s,\beta)$,
but of course it will depend on the system under consideration.  We will return
to this issue in Sec.\ \ref{SecCutoff}.

\begin{figure}[t]
\includegraphics[width=3in]{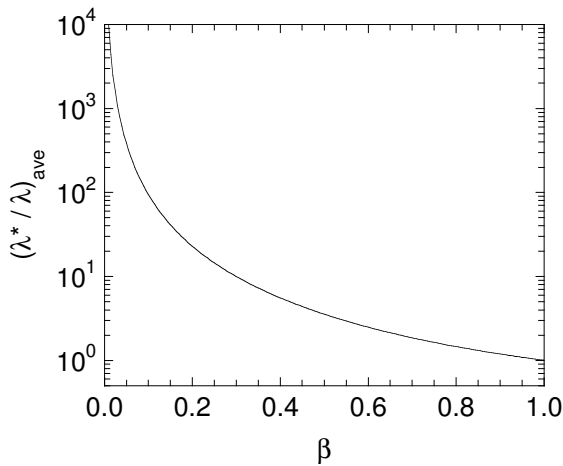}
\caption{Semilog plot of the average $(\lambda^*/\lambda)_{\rm ave}$ versus the
stretching exponent $\beta$ according to Eq.\ (\protect\ref{Eqsnave}) with $m =
1$.}
\label{Fig(1/lambda)ave}
\end{figure}

On the other hand, the $n = -1$ moment of $P(s,\beta)$, which is the average
$(1/s)_{\rm ave} = (\lambda^*/\lambda)_{\rm ave}$, is finite for $0 < \beta \leq
1$ and increases monotonically and rapidly with decreasing $\beta$, diverging at
$\beta = 0$.  In general, one has\cite{Lindsey1980}
\begin{equation}
\Big[\Big({\lambda^*\over\lambda}\Big)^m\Big]_{\rm ave} =  {\Gamma(m/\beta)\over
\beta \Gamma(m)}\ ,
\label{Eqsnave}
\end{equation}
where $m = -n$.  A semilog plot of $(\lambda^*/\lambda)_{\rm ave}$ versus $\beta$
for $0.01 \leq \beta \leq 1$ is shown in Fig.\ \ref{Fig(1/lambda)ave}. 

\subsection{Physical interpretations of $\lambda^*$ and $\beta$}

The parameter $\lambda^*$ in the stretched exponential
function (\ref{EqStretch}) is often referred to in the literature as some
undefined ``characteristic relaxation rate'' or as some undefined ``average''
relaxation rate.  As shown above, $\lambda^*$ is neither the average of $\lambda$
[which is infinite in the absence of a high-$s$ cutoff to
$P(s,\beta)$] nor the inverse of the average of $1/\lambda$.  Evidently
$\lambda^*$ is a characteristic property of $P(s,\beta)$ itself and not of its
moments.  Indeed, from the above discussion and the data in Fig.
\ref{FigPIntegral(0to1)}, we infer that the physical
interpretation of $\lambda^*$ is that $\lambda$ is
about equally likely to be less than $\lambda^*$ as it is to be greater (to
within $\sim \pm 20\%$).  From Fig.\ \ref{FigP(s)Median}, one sees
that $\lambda^*$ is within about 10\% of the median of $P(s,\beta)$ for $0.5 \leq
\beta \leq 1$.  For $\beta < 0.5$, the median strongly increases due to the long
high-$s$ tail to
$P(s,\beta)$, but the integral in Fig.\ \ref{FigPIntegral(0to1)} continues to
decrease slowly with decreasing $\beta$.

\begin{figure}[t]
\includegraphics[width=3in]{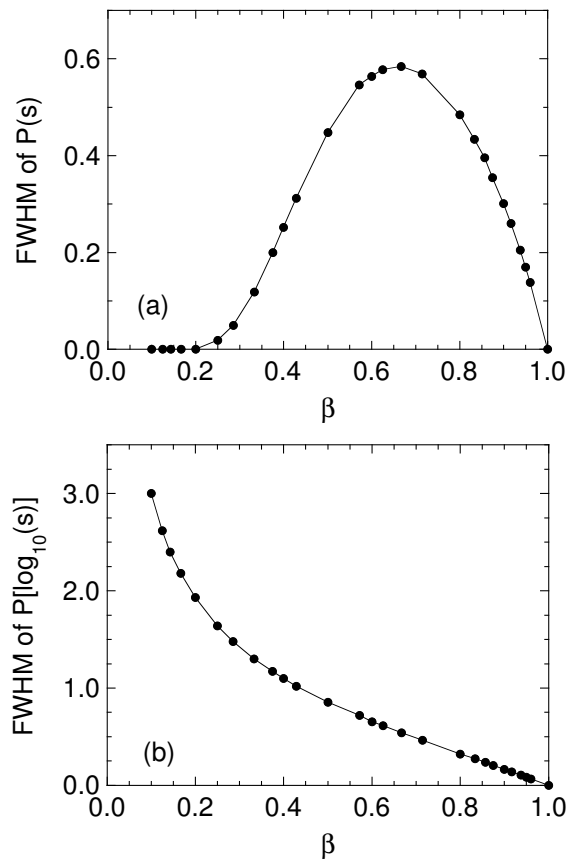}
\caption{Full width at half maximum peak value (FWHM) versus $\beta$ of the
probability distribution $P(s)$ on a linear $s$ scale (a) and of $P(\log_{10}(s))$
on a logarithmic $s$ scale (b).  See Figs.\ \protect\ref{FigP(s,beta)}(a) and
(b), respectively.  The solid curves are guides to the eye.}
\label{FigFWHM_of_P(s)}
\end{figure}

The stretching exponent $\beta$ is often cited as a
measure of the width $\Delta s$ of the distribution $P(s,\beta)$.  However, a
statistical definition of the width such as the rms width $\Delta s =
\sqrt{(s^2)_{\rm ave} - (s_{\rm ave})^2}$ is undefined for $P(s,\beta)$ since as
shown above both of the averages in the square root are infinite in the absence of
a high-$s$ cutoff to $P(s,\beta)$.  Another possibility is that $\beta$ is
a measure of the full width at half maximum ($\Delta s\equiv $ FWHM) of
$P(s,\beta)$.  From Fig.\ \ref{FigFWHM_of_P(s)}(a), the FWHM
initially increases as $\beta$ decreases below unity, but then decreases as
$\beta$ decreases further for $\beta \lesssim 0.65$.  Therefore, $\beta$ is a
multivalued function of FWHM, and hence correlation with the FWHM is not
a useful physical interpretation of $\beta$.  On a logarithmic $s$ scale, the FWHM
increases monotonically with decreasing $\beta$ as shown in Fig.\
\ref{FigFWHM_of_P(s)}(b), and appears to diverge as $\beta\to 0$.  Therefore
$\beta$ can be considered to be related to the \textit{logarithmic} FWHM of
$P(s)$, but not to the FWHM itself.

Therefore we seek a physical interpretation of $\beta$ that
is not explicitly tied to the width of $P(s,\beta)$ versus $s$.  We have seen
above that the position $s(P^{\rm max})$ of the peak in $P(s,\beta)$, plotted in
Fig.\
\ref{Figs^max(beta)} versus $\beta$, decreases monotonically with decreasing
$\beta$, and that below the peak, $P(s,\beta)$ decreases exponentially as $s$
decreases.  As discussed above, this demonstrates that $P(s,\beta)$ has a low-$s$
cutoff that decreases monotonically with decreasing $\beta$.  Thus a useful 
physical interpretation of $\beta$ is that $\beta$ is a measure of the intrinsic
small-relaxation-rate cutoff of $P(s,\beta)$.  To our knowledge, this
identification  and the above physical interpretation of $\lambda^*$ here and in
Ref.\
\onlinecite{Johnston2005} have not appeared elsewhere in the literature.

One can rewrite the stretched exponential function (\ref{EqStretch}) as
\begin{equation} 
n = n_0 \exp[-(t/\tau^*)^\beta]\ ,
\label{EqStretch2}
\end{equation}
where the relaxation time of a particular relaxing species in the system is $\tau
= 1/\lambda$ and the characteristic relaxation time of the probability
distribution is $\tau^* = 1/\lambda^*$, so $\tau/\tau^* = \lambda^*/\lambda$. 
From the above results, one obtains, for example, that the plot of
$(\lambda^*/\lambda)_{\rm ave}$ versus $\beta$ in Fig.\
\ref{Fig(1/lambda)ave} is the same as a plot of $\tau_{\rm ave}/\tau^*$
versus $\beta$.  From Fig.\ \ref{Fig(1/lambda)ave}, the
normalized average relaxation time $\tau_{\rm ave}/\tau^*$ is well-defined at
each $\beta$, increases monotonically and rapidly as $\beta$ decreases below
unity, and diverges as $\beta \to 0$.  Thus an additional physical
interpretation of $\beta$ is that $1/\beta$ is a measure of the average
relaxation time of the relaxing species in the system relative to the parameter
$\tau^* = 1/\lambda^*$.

\subsection{Influences of a large-$s$ cutoff to $P(s,\beta)$\label{SecCutoff}}

We have alluded above to the singular nature of the stretched exponential function
(\ref{EqStretch}) at time $t = 0$.  Associated with this singularity are infinite
averages $(s^n)_{\rm ave}$ with positive integer
$n$ for any fixed $\beta$ with $0 < \beta < 1$.  
Here we briefly discuss the
influences of a large-$s$ cutoff $s_{\rm cutoff}$ to $P(s,\beta)$ on 
$s_{\rm ave}$ and on the relaxation function that can be computed from the
modified $P(s,\beta, s_{\rm cutoff})$.  That relaxation function is no longer a
pure stretched exponential.  By definition, then,
\begin{equation}
P(s,\beta, s_{\rm cutoff}) = 0 \ \ \ {\rm for}\ s \geq s_{\rm cutoff}\ .
\label{EqCutoff}
\end{equation}

For illustrative purposes, we consider the probability
distribution for $\beta = 1/2$ because of the simple form of
$P(s,\beta = 1/2)$ in Eq.\ (\ref{EqP12}).  This $\beta$ value is also commonly
observed in real systems so the discussion here may have practical applications. 
The only influence of the cutoff on $P(s,\beta)$, beyond the cutoff condition
(\ref{EqCutoff}), is that $P(s,\beta)$ must be renormalized so that 
\begin{equation}
\int_0^{s_{\rm cutoff}} P(s,\beta, s_{\rm cutoff}) ds = 1\ .
\label{EqCutoff2}
\end{equation}
Applying conditions (\ref{EqCutoff}) and (\ref{EqCutoff2}) to $P(s,1/2)$ in Eq.\
(\ref{EqP12}) gives, for $s \leq s_{\rm cutoff}$,
\begin{equation}
P(s,1/2, s_{\rm cutoff}) = {\exp(-{1\over 4s})\over \sqrt{4\pi s^3}\ {\rm
erfc}({1\over 2\sqrt{s_{\rm cutoff}}})}\ ,
\label{EqP12cutoff}
\end{equation}
where ${\rm erfc}(z)$ is the complementary error function.  

\begin{figure}[t]
\includegraphics[width=3in]{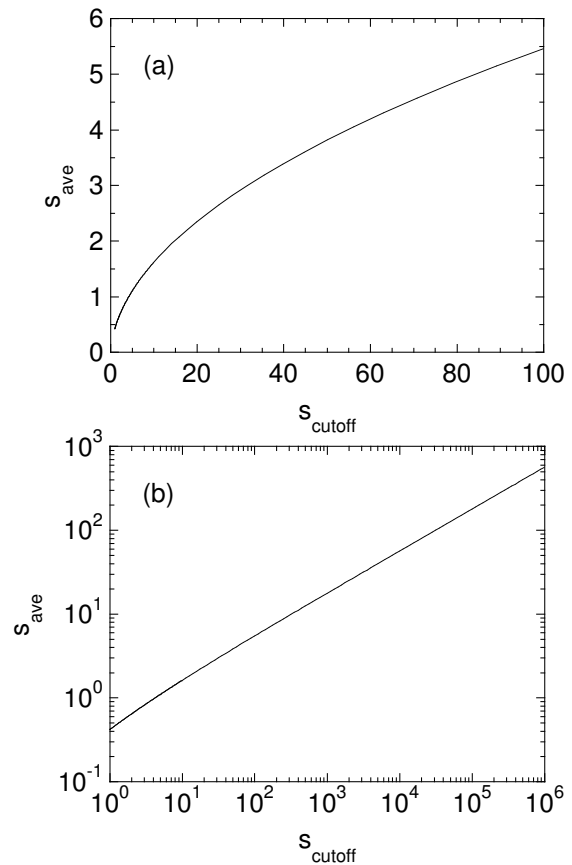}
\caption{Linear (a) and log-log (b) plots of $s_{\rm ave} = \lambda_{\rm
ave}/\lambda^*$ in Eq.\
(\protect\ref{EqSaveCutoff}) versus the large-$s$ cutoff $s_{\rm cutoff}$ to the
probability distribution $P(s,\beta = 1/2,s_{\rm cutoff})$  in Eq.\
(\protect\ref{EqP12cutoff}).}
\label{s_ave_vs_cutoff}
\end{figure}

The average of $s$ is $s_{\rm ave} = \int_0^{s_{\rm
cutoff}}sP(s,\beta,s_{\rm cutoff})ds$, yielding for $\beta = 1/2$
\begin{equation}
s_{\rm ave} = {\sqrt{s_{\rm cutoff}}\ \exp\big(-{1\over 4 s_{\rm
cutoff}}\big)\over
\sqrt{\pi}\ {\rm erfc}({1\over 2\sqrt{s_{\rm cutoff}}})}\ .
\label{EqSaveCutoff}
\end{equation}
Linear and log-log plots of $s_{\rm ave}$ versus $s_{\rm cutoff}$ are shown in
Figs.\ \ref{s_ave_vs_cutoff}(a) and \ref{s_ave_vs_cutoff}(b), respectively.  For
$s_{\rm cutoff}
\gg 1$, one can approximate Eq.\ (\ref{EqSaveCutoff}) as
\[
s_{\rm ave} = {1\over \pi} - {1\over 2} + \sqrt{{s_{\rm cutoff}\over \pi}}\ \ \ 
\ \ (s_{\rm cutoff} \gg 1)\ .
\]

\begin{figure}[t]
\includegraphics[width=3in]{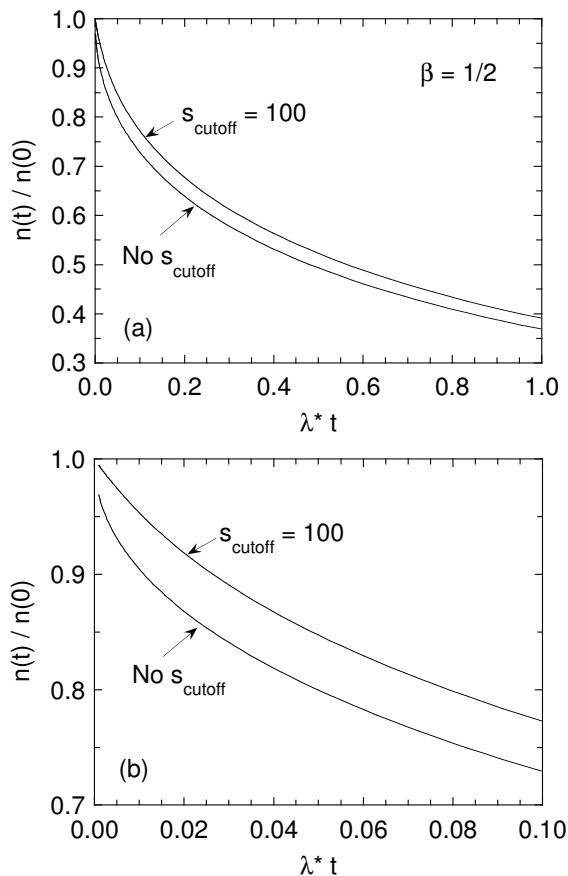}
\caption{(a) Comparison of the time $t$ dependent stretched exponential
relaxation with $\beta = 1/2$ without a cutoff to the probability distribution
and with a cutoff $s_{\rm cutoff} = 100$.  (b) Expanded plot of the data in (a)
near $t = 0$.  Note the initial linear dependence of $n$ on $t$ with the finite
cutoff $s_{\rm cutoff} = 100$.}
\label{n_vs_t}
\end{figure}

From Eq.\ (\ref{EqProb}), one obtains the time dependence of the relaxation of
the quantity $n(t)$ using
\begin{equation}  
{n(t)\over n_0} = \int_0^\infty P(s,\beta)e^{-s\lambda^* t}ds
\label{EqProb2}
\end{equation}
or, with a cutoff present,
\begin{equation}  
{n(t)\over n_0} = \int_0^{s_{\rm cutoff}} P(s,\beta, s_{\rm
cutoff})e^{-s\lambda^* t} ds\ .
\label{EqProb3}
\end{equation}
Shown in Fig.\ \ref{n_vs_t}(a) are plots of $n/n_0$ versus $t$ for
$\beta = 1/2$ and for $s_{\rm cutoff} = 100$ and $\infty$, where $P(s,1/2, 
s_{\rm cutoff})$ is given in Eq.\ (\ref{EqP12cutoff}).   Expanded plots for
$\lambda^*t \ll 1$ are shown in Fig.\ \ref{n_vs_t}(b). 
If there is no large-$s$ cutoff, then $s_{\rm ave} = \infty$ and the initial
slope of $n/n_0$ versus $t$ is $-\infty$ for $0 < \beta < 1$ as previously
discussed.  However, the presence of the cutoff gives a finite $s_{\rm ave}$ and
an initial linear decrease of $n(t)/n_0$ versus time as seen in Fig.\
\ref{n_vs_t}(b).  In that case, at sufficiently small $t << (s_{\rm
cutoff}\lambda^*)^{-1}$ one can Taylor expand the exponential in Eq.\
(\ref{EqProb3}) to get
$e^{-s\lambda^*t}\approx 1 - s\lambda^*t$.
Substituting this expression for the exponential into Eq.\ (\ref{EqProb3}) gives
\[
{n(t\to 0)\over n_0} = 1 - s_{\rm ave} \lambda^*t = 1 - \lambda_{\rm ave} t\ ,
\]
where we have used $s_{\rm ave} = \lambda_{\rm ave}/\lambda^*$.
The initial slope of $n/ n_0$ versus $t$ is then a true measure of the
negative of the average relaxation rate $\lambda_{\rm ave}$ in the system.  Every
real system must have a large-$s$ cutoff.  Experimentally, an important
resolution issue is being able to look at short enough times to be confident that
one is measuring the initial slope of the relaxation.  The length of time that
the initial linear relaxation is retained increases as
$s_{\rm cutoff}$ decreases.

\section{Summary}

The ubiquitous stretched exponential relaxation function (\ref{EqStretch}) has
been found to apply to the relaxation behavior of many different systems.  In
this paper we have examined some systematics of the probability distribution
$P(s,\beta)$, where $s = \lambda/\lambda^*$, that apply in the particular case
that the stretched exponential function arises from the global sum of exponential
decays of independently relaxing species with relaxation rates $\lambda$.

The functional
dependence of the peak position of $P(s)$ on $\beta$ has been determined.  This
peak position decreases monotonically with decreasing
$\beta$ and was characterized as a measure of the intrinsic low-$s$ cutoff
possessed by $P(s)$.  We therefore suggested that a physical interpretation of
$\beta$ is that $\beta$ is a measure of this intrinsic low-$s$ cutoff. 
Additionally, $1/\beta$ is a measure of the average relaxation time (not rate) of
the relaxing constituents of the system.  These interpretations contrast with a
common one that $\beta$ is a measure of the width in $s$ of $P(s,\beta)$, which
was shown to be not quantitatively useful.

We derived and discussed the $\beta$-dependence of the median of $P(s)$.  The
positive moments of $P(s)$, which are the averages $(s^n)_{\rm ave}$, are
infinite in the absence of a large-$s$ cutoff to $P(s,\beta)$, so $\lambda^*$ is
not related to any such average as often assumed.  We suggested instead that the
fundamental physical interpretation of $\lambda^*$ is that $\lambda$ is about
equally likely (to within $\sim \pm 20$\%) to be less than $\lambda^*$ as it is to
be greater.

The influence of a large-$s$ cutoff to
$P(s,\beta)$ on $s_{\rm ave} = \lambda_{\rm ave}/\lambda^*$ and on the
time dependence of the relaxation were investigated for the
illustrative case of
$\beta = 1/2$.  The cutoff leads to a finite average relaxation rate that
decreases as the cutoff decreases.  The cutoff also removes the infinite-slope
singularity in the stretched exponential relaxation at time $t = 0$ and replaces
it with an initial linear time dependence, the slope of which is the negative of
the average relaxation rate of the constituent relaxing species in the system. 

We note that experimental
time-dependent relaxation data can often be fitted by a discrete sum of
exponential relaxations, as well as by the stretched exponential function.  Thus
experimental resolution is important to distinguishing between such similar types
of fits.  Ultimately, the justification for using the stretched exponential
function to fit relaxation data, as opposed to some other function such as a
discrete sum of decaying exponentials, must reside in the physics of the system
under consideration.

\begin{acknowledgments} 

The author is grateful for the collaboration of and discussions with his
coauthors of Ref.\ \onlinecite{Johnston2005}. 
Ames Laboratory is operated for the U.S.\ Department of Energy by Iowa State
University under Contract No.\ W-7405-Eng-82. This work was supported by the
Director for Energy Research, Office of Basic Energy Sciences.
\end{acknowledgments}

\end{document}